# Analytical formulation for the bend-loss in single-ring hollow-core photonic crystal fibers

Michael H. Frosz,* Paul Roth, Mehmet C. Günendi, and Philip St.J. Russell

*Max Planck Institute for the Science of Light, Staudtstr. 2, D-91058 Erlangen, Germany*
*Corresponding author: michael.frosz@mpl.mpg.de*

Understanding bend-loss in single-ring hollow-core photonic crystal fibers is becoming of increasing importance as the fibers enter practical applications. While purely numerical approaches are useful, there is a need for a simpler analytical formalism that provides physical insight and can be directly used in the design of PCFs with low bend-loss. We show theoretically and experimentally that a wavelength-dependent critical bend radius exists below which the bend-loss reaches a maximum, and that this can be calculated from the structural parameters of a fiber using a simple analytical formula. This allows straightforward design of single-ring PCFs that are bend-insensitive for specified ranges of bend radius and wavelength. It also can be used to derive an expression for the bend radius that yields optimal higher-order mode suppression for a given fiber structure. © 2016 Chinese Laser Press
  OCIS codes: 060.2280, 060.2310, 060.4005, 060.5295.

## 1. INTRODUCTION

Hollow-core photonic crystal fibers (HC-PCFs) permit light to be guided over long distances in vacuum, gases, or liquids [1]. The lowest losses (<2 dB/km) have been reported in HC-PCFs that use a photonic bandgap to confine the light over a transmission window of ~100 nm [2, 3]. Other types of HC-PCF can achieve relatively low loss (<1 dB/m) over a broad transmission window (hundreds of nanometers) without having a photonic bandgap but instead relying on anti-resonant reflection from the periodic structure surrounding the core [4]; these include kagomé-style PCFs [5] and PCFs with a single ring of capillaries encircling the core [6]. Single-ring PCFs have recently been receiving increasing attention because of their relative simplicity and surprisingly low loss, enhanced recently by the discovery that higher-order modes can be efficiently suppressed when the ratio between the inner diameter $d$ of the cladding capillaries (anti-resonant elements or AREs) and the diameter $D$ of the core (defined as the minimum distance between two diametrically opposite AREs) is close to 0.68 [7]. Earlier experimental studies focused on fibers with $d/D$ much smaller than 0.68 [6, 8, 9]. As $d/D$ increases, however, the fiber becomes increasingly bend-sensitive. Although it is known that this occurs due to coupling between the core mode and the surrounding capillaries [10-12], no simple analytical expression has yet been reported that quantifies the bend-sensitivity for a given structure, although such an expression would permit easy optimization of the fiber design. Numerical solutions of Maxwell's equations yield loss values for specified values of bend-radius, structure and wavelength (e.g. [13, 14]), but provide only limited physical understanding of the underlying bend-loss mechanisms.

  We here derive, based on an intuitive physical picture, a simple analytical expression for the critical bend-radius above which bend-loss is negligible.

## 2. THEORY

Bend loss in fibers comes in two main forms. First, the transition bend-loss, which occurs as the radius of curvature is gradually reduced, and second the leakage loss of the eigenmode of the constant-curvature fiber [15]. In this paper we shall concentrate on the second of these.

  The analysis starts with an approximate expression for the effective phase index of the $LP_{pq}$-like mode in a circular glass capillary, first derived by Marcatili and Schmeltzer [16]:

$$n_{pq} = \sqrt{1 - \left(\frac{u_{pq}}{\pi}\frac{\lambda}{d_i}\right)^2} \quad (1)$$

where $u_{pq}$ is the $p$-th zero of the Bessel function $J_q$, $\lambda$ is the wavelength and $d_i$ is the inner diameter of the capillary ($D$ for the core and $d$ for the capillaries – see Fig. 1(a)).

  When the fiber is bent to radius of curvature $R$, the refractive index distribution becomes tilted (to first order) as follows [17]:

$$\hat{n}(x,y) = \left(1 + \frac{y}{R}\right)\bar{n}(x,y), \quad (2)$$

where $R$ is the bend-radius and the $(x, y)$ axes lie in the local transverse fiber plane, the $y$-axis pointing normal to the bend, the fiber axis being located at $y = 0$. The effective index of the $LP_{01}$ mode of a cladding capillary rotated at angle $\theta$ to the $y$-axis (see Fig. 1(a)) is then given by Eq. (2) with $y = [(d + D)/2]\cos\theta$.

  The index difference between an $LP_{pq}$ core mode and the $LP_{01}$ capillary mode takes the analytical form:

$$\Delta n_{01}^{pq} = \sqrt{1 - \left(\frac{u_{pq}}{\pi}\frac{\lambda}{D}\right)^2} - \sqrt{1 - \left(\frac{u_{01}}{\pi}\frac{\lambda}{d}\right)^2}\left(1 + \frac{d/D+1}{2R/D}\cos\theta\right). \quad (3)$$

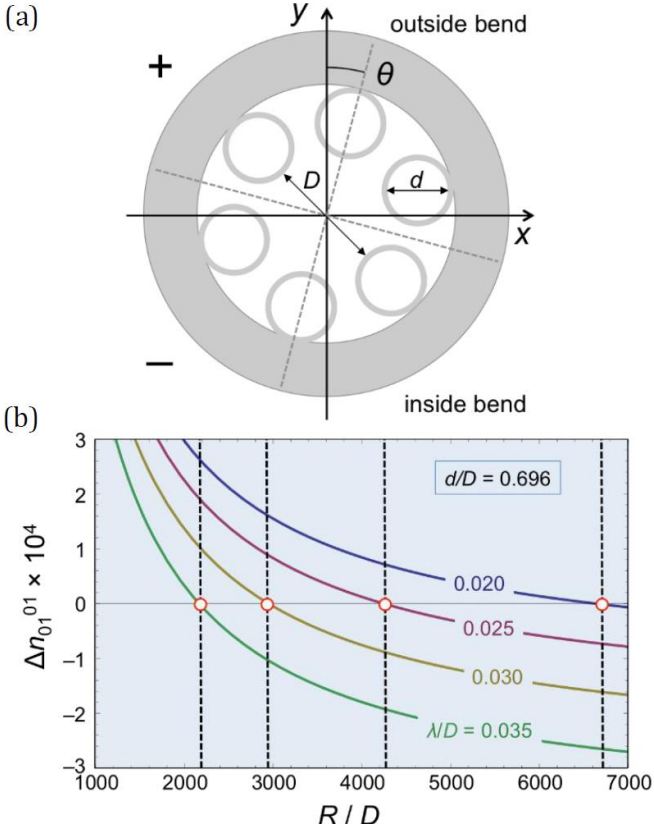

Fig. 1. (a) Sketch of the geometry of a single-ring hollow-core PCF, showing the local coordinate system. The inner diameter of the six capillaries is $d$ and the core diameter (the minimum distance between two diametrically opposite capillaries) is $D$. (b) Index difference $\Delta n_{01}^{01}$ between $LP_{01}$-like core and capillary modes, plotted against $R/D$ for $d/D = 0.696$ at four different values of $\lambda/D$.

As the radius of curvature falls, a critical value $R = R_{cr}^{pq}$ is reached when $\Delta n_{01}^{pq} = 0$ for $\theta = 0°$, i.e., when one of the cladding capillaries is at its maximum distance $(d + D)/2$ from the $x$-axis. Under these phase-matched conditions light couples strongly from the core mode to the outer capillary mode, where it leaks away rapidly into the surrounding solid glass sheath. If the fiber is carefully aligned so that $\theta = 30°$, the critical bend radius will be a factor $\sqrt{3}/2$ smaller. In a typical laboratory setting, however, $\theta$ will vary randomly so that the upper value of $R_{cr}^{pq}$ is likely to apply. Overall, this means that phase-matching between core and capillary modes can in principle occur for any bend-radius $R < R_{cr}^{pq}$. Starting with a straight fiber, therefore, the bend-loss will increase rapidly when the critical bend-radius is reached and the transmission will remain small for smaller values of $R$.

Figure 1(b) plots Eq. (3) for the case when $d/D = 0.696$, $p = 0$ and $q = 1$ (i.e., the $LP_{01}$ core mode, $u_{01} = 2.405$) for four different values of the scale parameter $\lambda/D$. This plot predicts, for example, a critical bend radius of ~17 cm for $D = 79$ µm, $d = 55$ µm and $\lambda = 2.8$ µm.

Using the approximation $\lambda/(\pi D) \ll 1$, Eq. (3) can be manipulated to yield an explicit expression for $R_{cr}^{01}$:

$$\frac{R_{cr}^{01}}{D} = \frac{D^2}{\lambda^2} \cdot \frac{\pi^2}{u_{01}^2} \frac{(d/D)^2}{1-d/D} \qquad (4)$$

which shows that for a given structure the radius of curvature at which bend-loss becomes significant scales with $D^3/\lambda^2$. This functional dependence is similar to that seen in standard step-index fiber, i.e., $R_{cr} \propto \rho^3/\lambda^2$ where $\rho$ is the core radius (see e.g. [18]). The expression in Eq. (4) is simple and has the advantage of clearly exposing the dependence of $R_{cr}^{01}$ on $D$, $\lambda$ and $d/D$. (Note that a more complicated analysis, based on similar physical assumptions, has been reported in connection with THz guidance in a single-ring polymer structure [10]).

Numerically simulated modal field (axial component of the Poynting vector) distributions for the structure mentioned above are plotted in Fig. 2 for three different values of $R$. At $R = 22$ cm, just above $R_{cr}^{01} = 17$ cm, the outermost capillary mode is only weakly excited and the loss is ~1.2 dB/m. At $R = 17$ cm, the bend-loss rises to ~21 dB/m and the outermost capillary lights up brightly. At $R = 8.6$ cm, the capillaries at $\theta = \pm 60°$ become resonant and the bend-loss drops to around 5 dB/m, with little difference between polarization states. Since in a typical experiment a fiber will start out straight and then gradually bend to smaller values of $R$, most of the bend loss will occur when $R$ has values close to $R_{cr}^{01}$. If $R$ remains constant along the whole fiber length, the loss can attain quite small values provided the core and capillary modes are phase-mismatched, which can occur even when $R$ is small.

The numerically calculated bend-losses versus bend-radii are plotted in Fig. 3 together with the critical bend-radius calculated from Eq. (4). It is clear (particularly in the linear plot) that Eq. (4) provides a good estimate for the bend-radius at which the bend-loss increases significantly. For both fibers, the bend-loss is of the order of 10 dB/m at the critical bend-radius. There is a small offset between the analytically calculated critical bend-radii and the peaks in bend-loss found from the numerical calculations, but this is to be expected in view of the simplifications made in deriving Eq. (4).

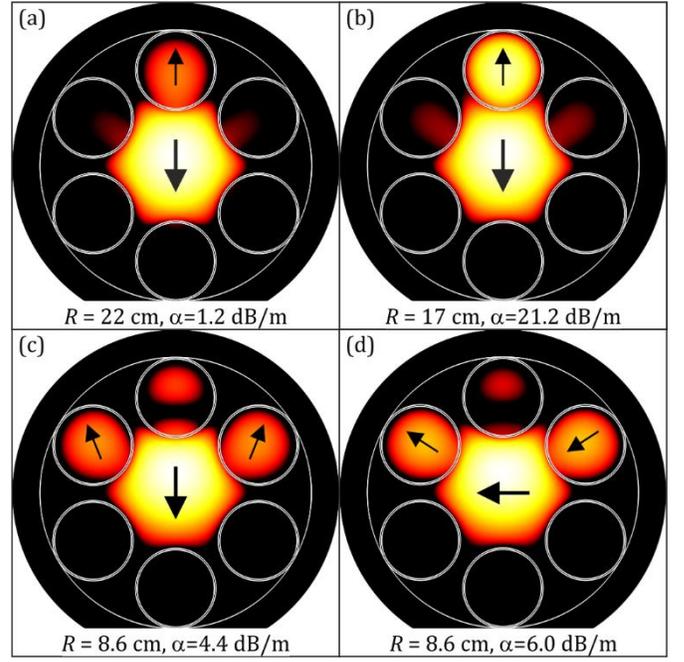

Fig. 2. Numerically calculated axial Poynting vector distributions and loss $\alpha$ of a single-ring PCF with $d = 55$ µm, $D = 79$ µm, $\lambda = 2.8$ µm and capillary wall thickness $t = 1.15$ µm, for (a) bend-radius slightly greater than $R_{cr}^{01} = 17.2$ cm, (b) close to $R_{cr}^{01}$ and (c,d) close to the radius of curvature that phase-matches the $LP_{01}$ core mode to capillaries at $\theta = \pm 60°$. The fiber is oriented as in Fig. 1(a) with $\theta = 0°$. The arrows indicate the polarization of the electric field.

## 3. EXPERIMENTS

To experimentally validate Eq. (4), the bend-loss in two different single-ring PCFs was measured by first recording the transmission spectra in few-m lengths of straight fiber. The fibers were then coiled around mandrels of varying radii (in steps of 1.25 cm) without disturbing the in- and out-coupling ends of the fibers, and the transmission spectra measured in each case. Subtracting the transmission spectrum (in dB) of the bent fiber from that of the straight fiber and dividing by the length of bent fiber section gave the bend-induced loss, which is plotted in Fig. 4. No care was taken to align the orientation of the fiber cross-section or the linear input polarization relative to the bend plane, because we are here only interested in the bend-sensitivity of these fibers under normal laboratory conditions when the fiber may be bent arbitrarily in any direction. The white curves in Fig. 4 plot $R_{cr}^{01}$ from Eq. (4) for both fibers. It is clear that the bend-loss remains relatively low for $R > R_{cr}^{01}$ but increases rapidly as $R_{cr}^{01}$ is approached, reaching a maximum as $R$ decreases beyond this point.

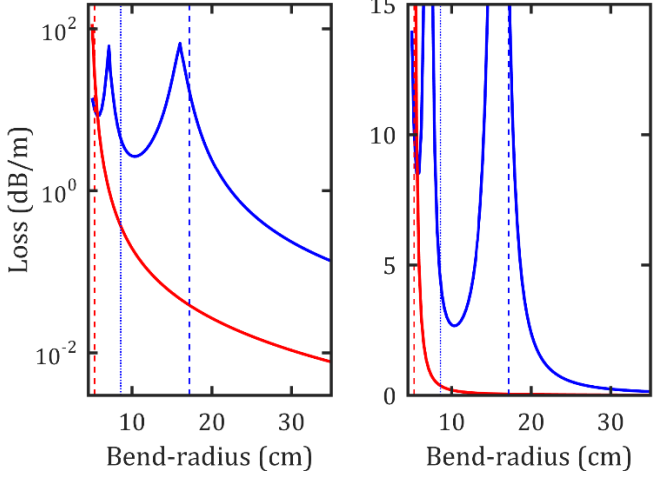

Fig. 3. Numerically calculated bend-losses for the fibers with $d/D = 55$ μm/79 μm (blue, $\lambda = 2.8$ μm) and $d/D = 22$ μm/36 μm (red, $\lambda = 1.2$ μm) plotted on semi-log scale (left) and linear scale (right). The thin dashed vertical lines show the corresponding analytical solutions for the critical bend-radius using Eq. (4); the dotted vertical line shows the critical bend-radius for phase-matching to the capillaries at $\theta = \pm 60°$. For each fiber, we show the loss of the modes polarized as shown in Fig. 2(a-c).

The gray-shaded rectangle in Fig. 4 (a) marks the wavelength range where there was no measurable transmission even in the straight fiber. This was caused by phase-matching to a resonance in the walls of the ring capillaries at wavelengths given by $\lambda_m = 2t(n^2-1)^{1/2}/m$, where $t$ is the capillary wall thickness, $n$ is the refractive index of the glass and $m$ is the order of the resonance [4, 19]. For the PCF in Fig. 4(a) $t = 1.15$ μm, yielding $\lambda_1 = 2.36$ μm. Figure 4(a) also shows that the fiber is highly bend-sensitive at wavelengths close to these resonances. For the PCF in Fig. 4(b) $\lambda_1 \sim 0.97$ μm ($t \sim 0.46$ μm), which lies outside the wavelength range considered here.

## 4. HIGHER-ORDER MODE SUPPRESSION

Moving on now to bend-loss for the LP$_{11}$–like core mode ($u_{11} = 3.832$) and applying the same approximations as used in deriving Eq. (4), we arrive at the result:

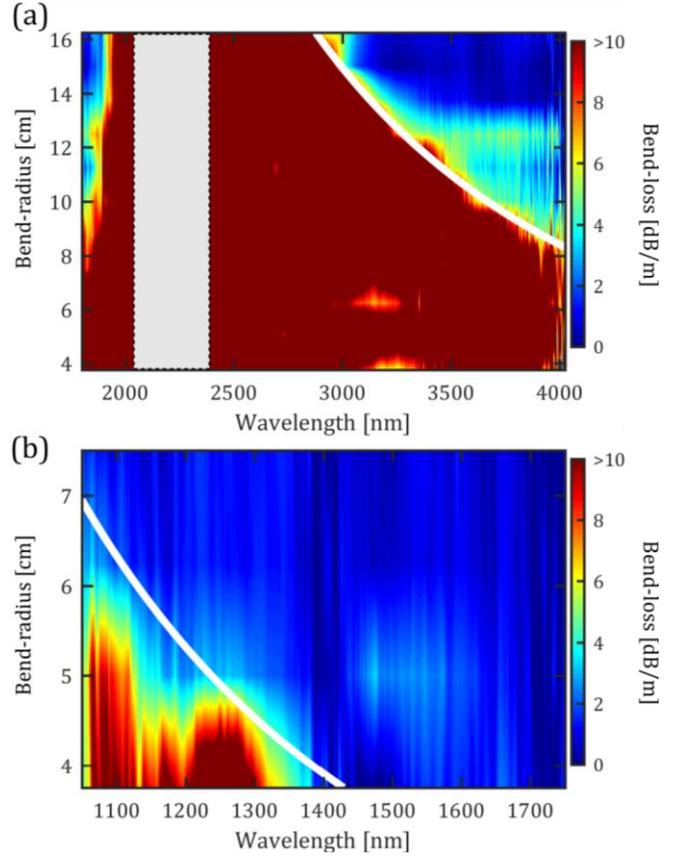

Fig. 4. Experimentally measured bend-loss in two different fibers with structural parameters (a) $d/D = 55/79$ and (b) $d/D = 22/36$. The bend radii were changed in steps of 1.25 cm and between these steps the colors are interpolated. The measured loss versus wavelength in (b) was smoothed with a moving average filter. The gray rectangle in (a) marks the region where the core mode phase-matches to a resonance in the walls of the capillaries, causing very high attenuation. The white curves plot Eq. (4) in each case.

$$\frac{R_{cr}^{11}}{D} = \frac{D^2}{\lambda^2} \cdot \frac{\pi^2(1+d/D)}{(u_{01}D/d)^2 - u_{11}^2}. \quad (5)$$

Under certain conditions, $\Delta n_{01}^{11}$ can be negative in the straight fiber in this case, which has the interesting consequence that light leaks away towards the *inside* of the bend for bend-radii less than $R_{cr}^{11}$, i.e., for $\theta = 180°$. This occurs when $d/D > u_{01}/u_{11} \approx 0.63$. As reported previously [7], efficient suppression of higher-order core modes in a straight single-ring PCF occurs when $d/D \approx u_{01}/u_{11}$. Equation (5) shows that higher-order core modes can be suppressed in cases when $d/D < 0.63$ (perhaps because of fabrication difficulties [14]) by bending to the correct radius of curvature.

## 5. CONCLUSION

We derived a simple analytical expression that accurately estimates the critical bend-radius for single-ring hollow core PCFs. Its predictions agree well with experimental measurements. The analysis may also be applied to structures with different numbers of ring capillaries, e.g., eight [12].